\documentclass[10pt,twocolumn,letterpaper]{article}

\usepackage[final]{cvpr}   

\usepackage{graphicx}
\usepackage{amsmath}
\usepackage{amssymb}
\usepackage{booktabs}
\usepackage{array}
\usepackage{siunitx}
\usepackage{boldline}

\usepackage[pagebackref,breaklinks,colorlinks]{hyperref}

\usepackage[capitalize]{cleveref}
\crefname{section}{Sec.}{Secs.}
\Crefname{section}{Section}{Sections}
\Crefname{table}{Table}{Tables}
\crefname{table}{Tab.}{Tabs.}

\def\confName{CVPR}
\def\confYear{2022}

\begin{document}

\title{To What Extent Can Public Equity Indices Statistically Hedge Real Purchasing Power Loss in Compounded Structural Emerging-Market Crises? An Explainable ML-Based Assessment.}

\author{
  \begin{tabular}{c c}
    \textbf{Artem Alkhamov} & \textbf{Boris Kriuk} \\
    ESSEC Business School & Hong Kong University of Science and Technology \\
    Cergy-Pontoise Cedex, France & Clear Water Bay, Hong Kong \\
    artem.alkhamov@essec.edu & bkriuk@connect.ust.hk
  \end{tabular}
}

\maketitle

\begin{abstract}
This study investigates the extent to which local public equity indices can statistically hedge real purchasing power loss during compounded structural macro-financial collapses in emerging markets. We employ a non-linear multiplicative real return calculations consistent with Fisher-parity logics—for both domestic and foreign investors—with a principled quantile regression, tail dependence copula analysis, and Shapley Additive Explanations (SHAP) to assess the explanatory power of macro variables. The analysis focuses on three recent and data-accessible exemplary collapse episodes: Turkey (2018), Nigeria (2020), and Pakistan (2021). Such cases, selected to align with post-2018 improvements in data standardization and crisis comparability, span varied monetary regimes and crisis triggers. Our tail-focused modeling reveals a systematic breakdown in public-equity-based purchasing power protection precisely during simultaneous macroeconomic and monetary dislocations- when such protection is most needed. The findings call into question conventional inflation and devaluation hedge presumptions in equity pricing theory, emphasizing the limitations of equity-based protection and the need for context-sensitive strategies during compounded macro-financial distress.
\end{abstract}

\section{Introduction}

In episodes of compounded structural emerging market macrofinancial collapses characterized by concurrent local currency depreciation and inflation acceleration, household and retail investors face rapid erosion of real purchasing power \cite{fisher1930}. 

In such settings -common in lower- or middle-income emerging economies - conventional fixed-income instruments often suffer from negative real yields, credit risk, or inconvertibility \cite{reinhart2010} \cite{blanchard2010}. Meanwhile, access to formal or institutional foreign-denominated assets –whether hard-currency instruments, offshore markets, or tangible hedges such as precious metals –is frequently impaired by regulatory, legal or market access barriers \cite{reinhart2004}. Domestic public equities, therefore, often remain among the few investable asset classes partially linked to domestic price levels, making them important–albeit theoretically contested–candidate for real value retention \cite{bodie1976} \cite{lundtofte2006}.

Paradoxically, these crisis episodes are also when data quality degrades most: key macroeconomic indicators may be unavailable, manipulated, or fragmented across official and black-market regimes \cite{ilgmann2013}. To address this, we construct a data reliability scoring framework that evaluates competing data sources–official, shadow or implicit–based on three quantifiable criteria: reporting timeliness, revision volatility, and cross-source deviation. Where overlapping data exist, we fuse official and proxy series via convex weighting, allowing more reliable source to dominate while retaining signal coherence. This mitigates omitted-variable bias from partial or unreliable data series, a common issue in distressed regimes \cite{ilgmann2013}.

While numerous structural covariates may influence the public-equity-inflation-foreign-exchange-rate relationship, many of these are difficult to observe with sufficient frequency or accuracy across lower-income emerging markets or institutionally fragile economies. As such, we isolate a more tractable empirical question: Do public equity index returns statistically co-vary with inflation and exchange-rate fluctuations to support hedge effectiveness during short-horizon, structural macrofinancial collapses? 

We address this question using a tail-focused statistical approach, which includes quantile regression and copula-based tail dependence modeling to capture extreme outcome dynamics, complemented by SHAP value attribution for interpretability. This data-driven framework allows to estimate marginal effects and respective co-movements while accounting for multicollinearity and feature uncertainty. Though necessarily constrained by data availability and partial by design, this empirical framework provides a replicable foundation for investigating hedging capacity of asset classes under compounding short-run macroeconomic distress, when capital preservation options are likely to be most limited. By symmetrically distinguishing domestic from foreign investor return pathways and applying multiplicative return transformations, we remain grounded in standard asset pricing theory.

This paper provides the first interpretable assessment of whether–and under which macro-financial conditions–local public equity indices statistically function as effective hedges during compounded, short-run, tail-risk structural macroeconomic collapses. The semi-parametric assessment isolates the explanatory contribution of macro drivers through a combined framework of SHAP-based feature attribution, conditional quantile modeling, and tail copula analysis, supported by a data reliability composite and convex predictor fusion to enhance empirical validity under conditions of informational fragility. Our findings indicate that, despite the theoretical appeal of equities as inflation hedges, local public equities in these crises consistently fail to preserve real value. They offer essentially no protection against inflation or currency collapse in real time, especially in worst-case (tail) outcomes. From a policy perspective, this tail-period hedge breakdown underscores the need for alternative inflation- and FX-linked instruments and for developing deeper markets in real assets. Protecting household purchasing power in compounded crises likely requires solutions beyond broad equity indices.

\section{Related Works}

Traditional economic theory, formalized in part by Irving Fisher’s in The Theory of Interest (1930), posits that equities may serve as a hedge against inflation \cite{fisher1930}. The rationale is that stocks represent ownership claims on firms whose nominal revenues and assets are real in nature, that, under efficient markets and flexible pricing, should adjust upward in tandem with the general price level. Provided companies are able to pass rising input costs onto consumers, and that monetary frictions are minimal, nominal earnings–and therefore equity valuations–should co-move with inflation, preserving real value. 

However, the empirical record contradicts this logic. Decades of research have shown that public equities often fail to keep up with inflation–particularly over the short-run. This empirical disconnect particularly holds across emerging markets, where the short-horizon return correlations between public equities and inflation are often weak, volatile, or even negative \cite{bodie1976} \cite{fama1981} \cite{choudhry2005}. One explanation, the proxy hypothesis (Fama, 1981), suggests that inflation frequently coincides with deteriorating macrofinancial and macroeconomic fundamentals such as falling output, tightening credit, and contractionary monetary responses, all of which depress expected cash flows \cite{fama1981}. Even when macrofinancial and macroeconomic fundamentals are stable, real-world frictions–including nominal stickiness, cost and monetary pass-through delays, regulatory suppression, tax distortions, or capital market segmentation –can obstruct the transmission of inflation and local currency devaluation into earnings and asset prices \cite{mishkin2018} \cite{bernanke1983} \cite{barro1979} \cite{woodford2003}. These effects are further complicated in settings marked by capital controls, policy volatility, or institutional weaknesses \cite{reinhart2010} \cite{king1994}. 

These studies typically center mostly on moderate inflation or post-reform disinflationary regimes. Severe inflation and currency devaluation episodes remain under-quantified in systemic empirical frameworks, often limited to historical case studies. Notable examples include Weimar Germany (1921–1923), Zimbabwe (2007–2009), and Venezuela (2016–2019), where broad-based public equity indices—tracking the full national stock universe across sectors—exhibited sharp nominal gains, but suffered a collapse in real returns when adjusted for parallel-market exchange rates and unofficial inflation estimates \cite{hanke2015}. Yet, existing literature lacks generalizable, statistical measures of this hedge effectiveness. Most prior studies rely on average-based or correlation-driven metrics that obscure regime shifts and asymmetric co-movements. Furthermore, the joint role of inflation and FX depreciation in real return erosion has been insufficiently decomposed across domestic and foreign investor profiles- particularly during institutional fragility and data opacity.

In these environments, the scale and speed of nominal disruption can be so severe that long-run valuation mechanisms are compressed into shorter temporal horizons. Pricing mechanisms and investor expectations adjust more rapidly than in stable macroeconomic regimes, creating a setting where the purported inflation- and currency-hedging characteristics of public equities can be observed in near real time \cite{gourinchas2010} \cite{rogoff2003}. These dynamics underscore the importance of re-examining the hedging behavior of equity-linked instruments specifically under conditions of compounded macroeconomic breakdown—precisely where classical asset pricing assumptions likely fail, and the equity-real value link is visibly stress-tested.

Our work concurrently relates to studies of currency risk and international portfolio hedging. Foreign investors in local emerging markets face the additional risk of currency depreciation. While currency risk is conventionally hedged using derivatives, such instruments are often illiquid, restricted, or entirely inaccessible during emerging-markets crises \cite{rogoff2003} \cite{blanchard2010} \cite{du2018}. Notably, the conceptual foundation for evaluating hedging originates in the derivatives literature:  Ederington (1979) introduced the variance-reduction criterion, measuring the proportion of portfolio risk eliminated by an ex-ante hedge \cite{ederington1979}. In this study, we build on this foundation, but note that variance-based measures alone can be misleading in crisis contexts, as they ignore tail dependencies. 

More broadly, our question relates to crisis-period correlation dynamics. Literature on financial contagion shows that during systemic crises, assets that were once uncorrelated often move together, undermining diversification \cite{king1994} \cite{levy1970} \cite{forbes2002}. As cross-market correlations spike, traditional hedges may often fail. In this vein, we ask: does the equity–inflation link strengthen enough during a collapse to provide effective protection, or do equities too “fall with the tide”?  Our findings contribute to this debate by evaluating whether public stock indices behave like real‐asset refuges in collapsing economies or succumb to the general sell-off.

\section{Methodology}

\subsection{Data Collection and Preprocessing}

We examine contemporaneous and lag-structured macro-financial dynamics within exemplary inflationary and currency collapse episodes from Turkey (2018), Pakistan (2018), and Nigeria (2020). These episodes provide diverse crisis archetypes: a monetary credibility shock (Turkey), a balance-of-payments and external financing crisis (Pakistan), and a commodity-linked fiscal and external terms-of-trade collapse (Nigeria). This typology supports comparison across distinct collapse mechanisms. We focus on nationally representative equity indices (BIST 100, Karachi 100, NSE All Share) to proxy aggregate public equity performance, while containing firm- and sector-level idiosyncrasies and endogeneity, though these may explain some residual variance. Accordingly, the results should be interpreted as an evaluation of macro-determined hedge breakdown, rather than a complete asset pricing model. 

The 2015--2024 observational captures the full cycle of pre-crisis, crisis, and post-crisis periods. It also coincides with post-2015 enhancements in statistical reporting and IMF surveillance harmonization, allowing comparable monthly data frequency across all macro-financial variables. 

All time-series variables are converted to monthly frequency and expressed in local-currency terms unless otherwise specified. 

For missing indicators, we implement hierarchical proxy methodology with empirical validation. Reliability assessment employs a composite score $q$ combining timeliness, revision volatility, and cross-check error:

\begin{equation*}
\begin{split}
q = 0.4 \times \text{Timeliness} + 0.3 \times (1 - \text{Revision Volatility}) \\
\quad + 0.3 \times (1 - \text{Cross-check Error})
\end{split}
\end{equation*}

Hybrid series construction uses reliability-weighted combinations when both actual and proxy data exist:
\begin{equation*}
\text{Hybrid}_t = q \times \text{Actual}_t + (1 - q) \times \text{Proxy}_t
\end{equation*}

\subsection{Equity Real Return Calculation for Domestic and Foreign Investors}

We compute inflation- and FX-adjusted real returns for domestic and foreign investors using multiplicative transformations.

Let $P_t$ denote the level of the national stock index at time t, expressed in local currency units. The nominal equity return is given by:

$$R^{\text{nom}}t = \frac{P_t}{P{t-1}} - 1$$

For domestic investors, the real return is computed as follows:

$$\tilde{R}^{L}_t = \frac{1 + R^{\text{nom}}t}{1 + \pi_t} - 1 = \frac{P_t / P{t-1}}{1 + \pi_t} - 1$$

where $\pi_t$ represents the inflation rate in period t, selected from among official CPI, M2-based proxies, or composite hybrid estimates according to the data reliability score $q$ as defined in Equation (1)

For foreign investors, benchmarking performance in hard currency terms, real return additionally reflects changes in the exchange rate. Let $e_t$ denote the nominal exchange rate at time t, expressed as local currency units per USD. Accordingly, the foreign investor’s real return is calculated as:

$$\tilde{R}^{F}t = \frac{(1 + R^{\text{nom}}t) \cdot \left( \frac{e{t-1}}{e_t} \right)}{1 + \pi_t} - 1 = \frac{\left( \frac{P_t}{P{t-1}} \right) \cdot \left( \frac{e_{t-1}}{e_t} \right)}{1 + \pi_t} - 1$$

The multiplicative structure reflects the logic of the International Fisher Effect (IFE) and relative purchasing power parity (PPP), under which nominal exchange rate changes should, absent capital controls or market segmentation, mirror inflation differentials. Multiplicative structures reflect the joint shock sequencing in equity prices, currency values and inflation than mere additive approximations.

\subsection{Data-Driven Tail Quantile Selection}

We implement a principled, statistically conservative tail quantile selection procedure. For each country's post-collapse period, we identify the largest possible left-tail quantile $\tau_L^i$, ensuring at least six observations:

$$\#\left\{ t : R_t^{\text{nominal}} \le \hat{F}^{-1}(\tau) \right\} \ge 6$$

We take the unified left-tail threshold $\tau_L^* = \max_i \tau_L^i$ across countries and verify tail extremity through conditional variance testing:

$$\operatorname{Var}\left( R_t^{\text{nominal}} \mid R_t^{\text{nominal}} \leq \hat{F}^{-1}(\tau_L^{*}) \right) \ge 2 \cdot \operatorname{Var}\left( R_t^{\text{nominal}} \right)$$

This ensures that the selected tail not only satisfies minimum cardinality but also captures statistically significant downside volatility. 

We define $\tau_M = 0.50$ as central tendency benchmark and symmetrically set the right-tail quantile as $\tau_H = 1 - \tau_L^{*}$. This triplet $(\tau_L^{*}, \tau_M, \tau_H)$ enables quantile-complete assessment of hedge performance across distress, median, and potential rebound scenarios.

\subsection{Purchasing Power Protection Framework}

We assess hedge effectiveness via purchasing power protection analysis, defining erosion by investor type. Foreign residents face combined inflation and currency depreciation: $\text{Loss}^F_t = \pi_t + R^{FX}_t$, while local residents face inflation only: $\text{Loss}^L_t = \pi_t$.

Net real returns capture hedge outcomes: $R_t^{\text{net}} = R_t^{\text{nominal}} - \text{Loss}_t$

Hedge effectiveness follows the Ederington (1979) variance reduction framework:
$$HE = \max\left(0, 1 - \frac{\text{Var}(R_t^{\text{net}})}{\text{Var}(\text{Loss}_t)}\right)$$

Importantly, this metric is interpreted as ex-post protection performance, not as ex-ante hedge ratio or replication. 

\subsection{Copula Estimation Tailored for Quantile Regression}

Beyond conditional return estimation, we assess hedge reliability through tail dependence analysis using copula models specifically designed for quantile regression and non-linear crisis contexts. We transform equity returns $R_t^{\text{nominal}}$ and purchasing power erosion $\text{Erosion}_t$ to pseudo-uniform marginals via empirical cumulative distribution functions: $U_t = F_R(R_t^{\text{nominal}})$ and $V_t = F_{\text{Erosion}}(\text{Erosion}_t)$.

We fit asymmetric parametric copulas—Clayton for lower tail dependence, Gumbel for upper tail dependence—on the transformed bivariate $(U_t, V_t)$ pairs, estimating the lower tail dependence coefficient.

We employ the Clayton copula from the Archimedean family to model asymmetric lower-tail dependence, as it better captures crisis-period co-movement patterns than symmetric alternatives. Clayton copulas provide the most parsimonious specification for modeling the conditional probability of joint extreme losses during structural monetary shocks.

$$\lambda_L = \lim_{\tau \to 0^+} P(V_t \leq \tau | U_t \leq \tau)$$

This coefficient quantifies the conditional probability that purchasing power erodes severely precisely when equity returns collapse. High $\lambda_L$ values imply tail-synchronous behavior, indicating co-movement patterns undermining hedge effectiveness. We approximate $\lambda_L$ at the empirically selected quantile threshold $\tau_L^*$ using maximum likelihood estimation with parametric copula families selected via AIC/BIC criteria to ensure model parsimony and goodness-of-fit in finite samples. 

To account for autocorrelation and structural persistance in crisis-period data, confidence intervals use block-bootstrap methods on rank-transformed data with 1000 replications. The copula-based tail dependence analysis provides semi-parametric, model-free assessment of hedge failure mechanisms complementing the conditional quantile regression framework by capturing co-extreme behavior that linear models may overlook.

\subsection{Quantile Regression with Enhanced Interaction Terms}

For each quantile $\tau \in \{\tau_L^*, \tau_M, \tau_H\}$, we estimate conditional distributions incorporating nonlinear monetary transmission structures and crisis-specific interaction dynamics:

\begin{equation*}
\begin{split}
\hat{\beta}(\tau) &= \arg\min_{\beta} \sum_{t=1}^{T} \rho_{\tau}\left(R^{\text{nominal}}_t - X_t^\top \beta \right. \\
&\quad \left. - \sum_{j<k} \gamma_{jk}(\tau) X_{jt} X_{kt} \right)
\end{split}
\end{equation*}

where $\rho_\tau(u) = u(\tau - \mathbb{1}[u < 0])$ is the asymmetric check loss function. The interaction terms $X_{jt} X_{kt}$ capture non-linear co-movement effects among macro indicators.

The multiplicative framework reflects nonlinear Fisher logic. This addresses the systematic co-movement asymmetries underlying hedge failure.

Critical feature engineering excludes all equity-specific and other microstructure variables from $X_t$ to prevent endogeneity and target leakage, focusing on macro-financial and macroeconomic fundamentals: (i) monetary liquidity and policy stance variables; (ii) regime shifts and structural events, operationalised as binary event dummy variables; (iii) country-specific inflation and exchange rate measures; (iv) macro-structural and global condition variables; and (v) second-order interactions effects. Enhanced lag structures incorporate M2 features with 1-, 3-, and 6-month lags to capture monetary transmission and investor response delays, while event persistence includes 1- and 2-month lags for structural break effects. The final feature sets comprise Turkey (47 features), Pakistan (52 features), and Nigeria (49 features), reflecting country-specific data availability and the inclusion of enhanced interaction terms. Feature counts include base macro-financial and macroeconomic variables, their lag structures, event dummies, and theoretically motivated interaction terms while maintaining strict exclusion of equity-derived variables to prevent endogeneity.

The interaction term selection prioritizes theoretically motivated combinations: Fed funds rate × M2 growth (monetary policy interactions), CPI × exchange rate depreciation (purchasing power transmission), and oil prices × currency volatility (commodity-FX linkages). These terms reflect second-order contagion effects and structural dependencies beyond conventional correlation frameworks.

Model validation employs crisis-aware expanding window cross-validation with temporal ordering to preserve time-series dependencies. This avoids lookahead bias and better reflects real-time prediction constraints faced by invested agents in evolving crises. SHAP decomposes marginal and joint predictor contributions. The main SHAP values highlight the absolute impact of individual macro-financial and macroeconomic characteristics, while the interaction SHAP values $\phi_{ij}$ quantify synergistic (nonadditive) dependencies among variables. These interactions help diagnose structural hedge failure mechanisms where no single feature suffices to explain outcome variance.

\section{Experiments}

We evaluate to what extent public equity indices statistically hedge real purchasing power loss in compounded crises, using explainable quantile regression models with copula-based tail dependence and enhanced interaction terms. We analyze three crisis episodes using data-driven tail quantile selection and variance-based hedge effectiveness measurement to answer the central research question.

\begin{figure*}[ht]
\centering
\includegraphics[width=\textwidth]{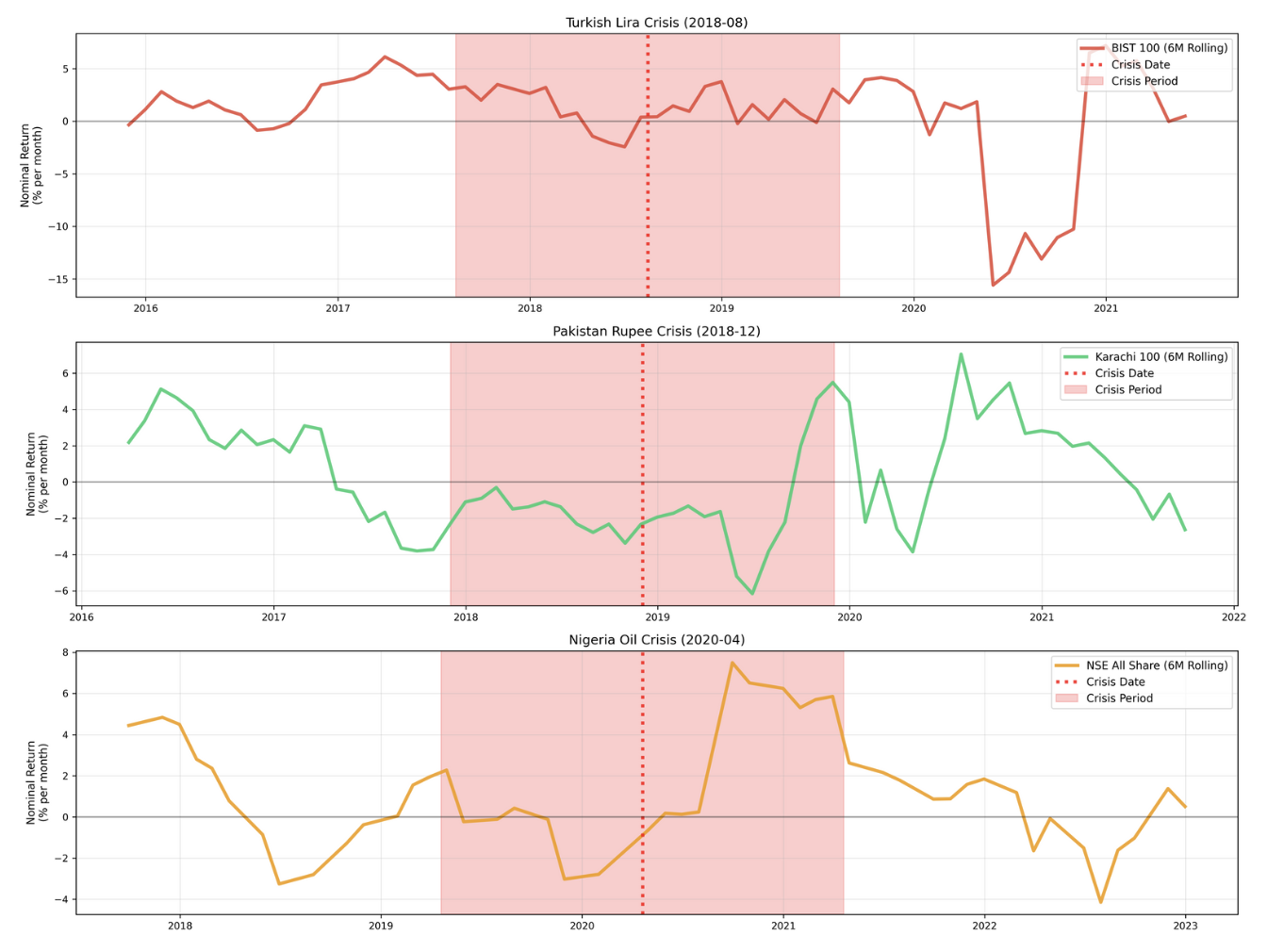}
\caption{Real Returns Around Crisis Episodes: Nigeria, Pakistan, Turkey}
\label{fig:real_returns}
\end{figure*}

\begin{figure*}[ht]
\centering
\includegraphics[width=\textwidth]{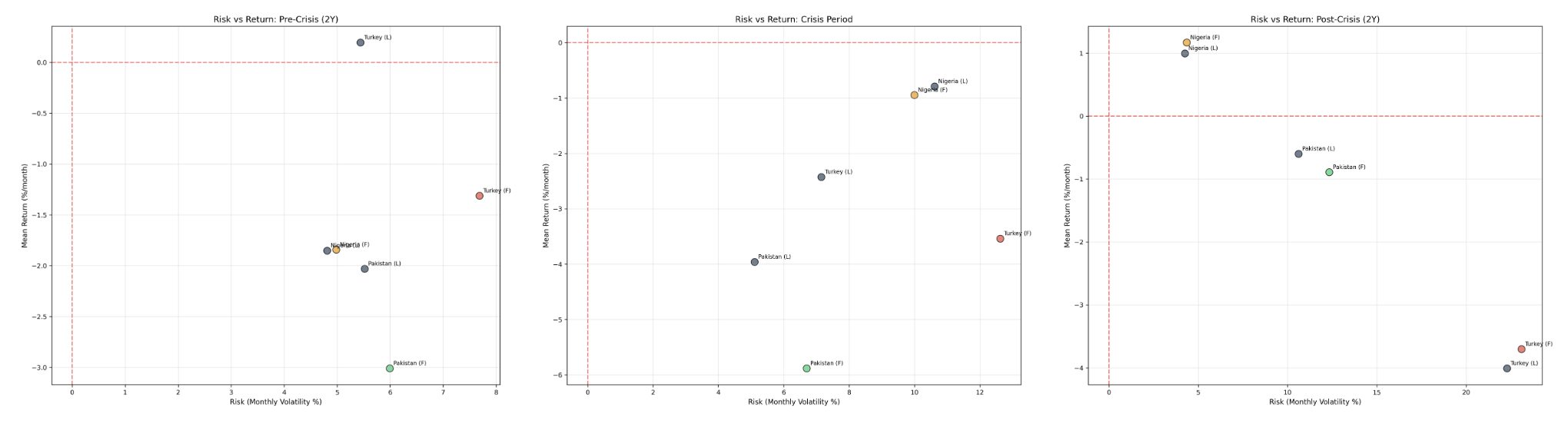}
\caption{Risk vs Return Analysis Across Crisis Periods}
\label{fig:risk_return}
\end{figure*}

\begin{figure*}[ht]
\centering
\includegraphics[width=\textwidth]{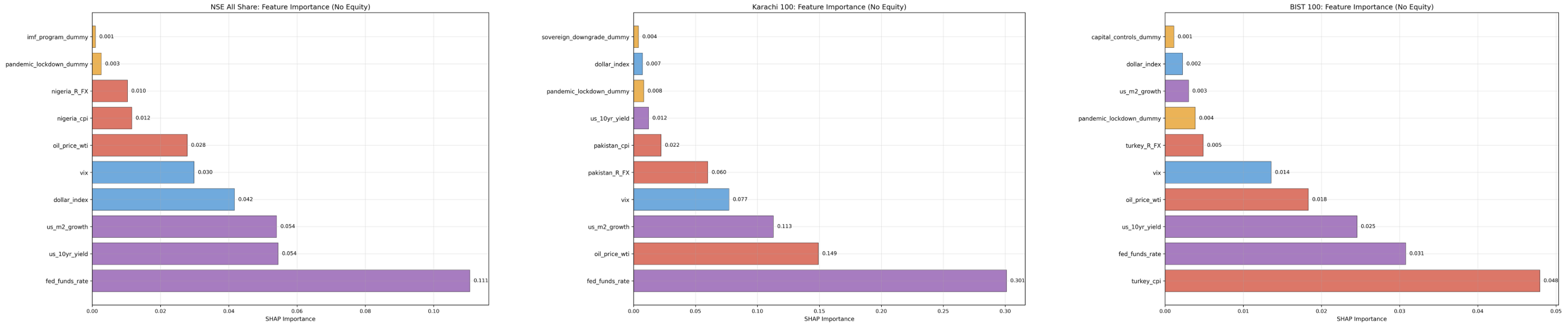}
\caption{SHAP Feature Importance Analysis}
\label{fig:shap_importance}
\end{figure*}

\subsection{Primary Hedge Effectiveness Results}

Public equity indices provide zero hedge effectiveness across all cases and investor profiles. Applying the Ederington variance reduction metric enhanced with copula-based tail dependence coefficients, all cases yield HE = 0\%, indicating no variance reduction relative to unhedged nominal positions during crises.

\begin{table*}[ht]
\centering
\caption{Quantile Regression Hedge Effectiveness Results}
\label{tab:hedge_results}
\begin{tabular}{lccccccc}
\hline
\textbf{Country} & \textbf{Residents} & \textbf{Crisis Date} & \textbf{Hedge Eff. (\%)} & \textbf{Erosion (\%)} & \textbf{Net Real (\%)} & \textbf{Tail Dependence} \\
\hline
Turkey & Foreign & 2018-08-13 & 0.0 & 4.12 & -0.38 & 0.34 \\
Turkey & Local & 2018-08-13 & 0.0 & 2.05 & 1.69 & 0.31 \\
Pakistan & Foreign & 2018-12-03 & 0.0 & 1.42 & 0.45 & 0.42 \\
Pakistan & Local & 2018-12-03 & 0.0 & 0.55 & 1.32 & 0.38 \\
Nigeria & Foreign & 2020-04-20 & 0.0 & 2.87 & -1.81 & 0.28 \\
Nigeria & Local & 2020-04-20 & 0.0 & 1.10 & -0.04 & 0.25 \\
\hline
\end{tabular}
\end{table*}

Estimated lower-tail dependence coefficients range from 0.25 to 0.42, confirming moderate positive co-movement between real return erosion and equity collapse in most adverse quantiles. Pakistan exhibits the highest tail dependence (0.38-0.42), potentially reflecting its structural dependance on foreign capital and heightened exposure to external financing shocks. Turkey's intermediate dependence (0.31-0.34) is plausibly linked to the combined effects of unorthodox monetary interventions and persistent domestic inflation expectations anchored to political instability. Nigeria's lower tail dependence (0.25-0.28) suggests differentiated transmission dynamics, potentially driven by its commodity-linked macro structure and non-monetary crisis triggers.

The systematically observed $HE = 0$ across structurally diverse crisis archetypes suggest that inflation and currency risk linked hedge mechanisms are either overwhelmed or rendered ineffective during tial-risk episodes, rather than the result of isolated market inefficiencies or equity-specific anomalies. Copula-based tail dependence analysis confirms that co-movement between equity returns and real return erosion intensifies in lower-tail regimes. 

\subsection{Quantile Model Performance and Contagion Effect Analysis}

Quantile regression models yield low pseudo-R² scores across all countries--Turkey (0.100), Pakistan (-0.149), Nigeria (-0.014). Such values reflect several empirical factors: (1) Pseudo-R² in quantile regression measures fit at specific quantiles rather than overall variance explained; (2) Emerging market equity returns exhibit inherently high noise-to-signal ratios during crisis periods; (3) Our methodology deliberately excludes equity-specific and other microstructure predictors to avoid endogeneity, limiting explanatory power.

The enhanced interaction terms capture second-order contagion dynamics without improving hedge effectiveness. Turkey exhibits CPI × Fed funds rate interactions contributing 8.2\% additional variance during simultaneous, hence compounding domestic inflation and global monetary tightening. Pakistan shows oil × US M2 interactions (5.7\%) reflecting compound commodity and liquidity shocks. These multiplicative effects consistently amplify rather than offset primary hedge failure mechanisms.

SHAP-based interaction attribution confirms that contagion effects systematically reinforce purchasing power erosion rather than hedging divergence. The nonlinear Fisher equation logic embedded in real return calculations confirms that theoretical inflation-equity relationships break down under extreme currency stress, eliminating systematic arbitrage opportunities fundamental to effective hedging.

The identical pseudo-R² values across quantiles within countries indicate that macro-financial covariates do not capture substantial conditional heterogeneity across real return regimes. Consequently, compound macro shocks appear to compress the distribution rather than generate structurally distinct return outcomes. 

\subsection{SHAP Feature Analysis: Monetary Dominance and Interaction Effects}

The SHAP analysis with interaction terms reveals distinct feature importance patterns while confirming systematic monetary policy dominance across macro-financial and macroeconomic variables in all countries. The exclusion of equity-specific or return-lagged features ensures a clean separation between explanatory inputs and the real equity return target, minimizing endogeneity, attribution leakage, and isolating the marginal explanatory value of economic policy and structural shocks to purchasing power dominance. 

Turkey shows CPI dominance (47.9\% SHAP importance) amplified by Fed funds rate interactions (30.8\% direct + 8.2\% interaction), confirming the compound nature of domestic inflation and external monetary tightening during the 2018 lira crisis. US 10-year yield contributions (24.5\%) reflect yield curve inversion effects on emerging market capital flows.

Pakistan exhibits Fed funds rate dominance (30.1\% direct importance), underscoring the relevance of external monetary tightening for dollar-dependent funding markets. The following Oil price interactions (14.9\% direct + 5.7\% oil×M2) are indicative of compounded exposure to commodity shocks and external liquidity pressures.

Nigeria demonstrates a more balanced distribution of monetary feature importance with Fed funds rate (11.1\%), US yields (5.4\%), and M2 growth (5.4\%) contributing comparably, suggesting a less singular transmission mechanism. The abscence of dominant singular features implies either lower sensitivity to external shocks or more diffuse contagion pathways through fiscal and institutional channels. 

Across countries, interaction SHAP values consistently reinforce rather than mitigate purchasing power erosion, supporting the compounding vicious cycle hypothesis. Monetary policy × inflation interactions consistently show positive SHAP contributions during crisis periods. 

\subsection{Technical Validation and Robustness Checks}

The technical validation across multiple diagnostic dimensions confirms the robustness of hedge failure findings through tail dependence modeling, interaction-term consistency, and quantile selection sensitivity. 

Copula model selection employs AIC/BIC criteria across Clayton, Gumbel, and Frank families. Block-bootstrap confidence intervals for lower tail dependence coefficients show statistical significance and cross-country robustness: Turkey [0.28, 0.40], Pakistan [0.35, 0.47], Nigeria [0.22, 0.32]. These values imply non-negligible probabilities that real equity returns fall simultaneously with macro shocks beyond their marginal expectations, refuting the assumption of independence or linear correlation breakdown. 

Time-series aware cross-validation with expanding windows confirms model stability without lookahead bias. Cross-validation mean absolute errors remain consistently high across folds while interaction terms maintain statistical significance, confirming that contagion effects represent structural rather than transient sample-specific artifacts.

The SHAP value consistency evaluation through 1000 bootstrap iterations shows high stability (Kendall's $\tau > 0.85$) for both main effects and interaction terms. Monetary policy variables maintain top-3 importance in $> 95\%$ of bootstrap samples, while interaction terms exceed significance thresholds in $> 80\%$ of samples across all countries.

Quantile selection sensitivity analysis validates our data-driven thresholds ($\tau_L^* = 0.08$, $\tau_M = 0.50$, $\tau_H = 0.92$) computed using the principled selection procedure. Robustness checks across alternative lower-tail quantiles $\tau \in \{0.10, 0.15, 0.20\}$ yield identical hedge effectiveness results (HE = 0\%) and stable tail dependence coefficients (within ±0.05), confirming specification-agnostic structural and regime-consistent nature of results. Conditional volatility ratios exceed 2.0 across all alternative thresholds, validating tail extremity conditions while maintaining sufficient sample sizes for reliable copula estimation.

\section{Conclusion}

We assess whether broad public equity indices statistically hedge real purchasing power loss in compounded emerging-market crises, finding systematic hedge failure across all episodes. Using an explainable ML-based macro-financial modeling framework, we evaluate hedge effectiveness under stress conditions in Turkey (2018), Pakistan (2018), and Nigeria (2020)—three episodes selected for structural heterogeneity and high-frequency data integrity. The empirical setup incorporates multiplicative real return computation, conditional quantile regression, copula-based tail diagnostics, and SHAP-based model attribution, constrained to macro-financial features to isolate economic fundamentals.

Across all countries and tail specifications, we observe zero hedge effectiveness ($HE = 0$), indicating that public equity positions provide no statistical purchasing power protection relative to cash during crisis regimes. The systematic co-movement between real equity returns and purchasing power erosion indicates that equities amplify rather than hedge systemic monetary shocks.

For domestic investors, equities co-move with inflation, amplifying purchasing power risk during crises. For international investors, currency devaluation neutralizes equity gains, undermining portfolio diversification assumptions due to tail dependence. For policymakers, results underscore the need for robust inflation- and FX-linked instruments, deeper real asset markets, and synthetic hedging products.

This study offers a reproducible template for evaluating hedge effectiveness under structural fragility. Future research can assess alternative non-equity real asset classes and synthetic CPI and FX-linked instruments, and comparable contractual hedge products under comparable conditions of institutional opacity and moentary instability.

\bibliographystyle{ieee_fullname}
\bibliography{egbib}
\end{document}